\documentstyle[12pt]{article} 
\newcommand{\be}{\begin{equation}}
\newcommand{\nn}{\nonumber}
\newcommand{\bea}{\begin{eqnarray}}
\newcommand{\eea}{\end{eqnarray}}
\newcommand{\ba}{\begin{array}}
\newcommand{\ea}{\end{array}}
\newcommand{\ee}{\end{equation}}

\expandafter\ifx\csname mathbbm\endcsname\relax

\else

\fi \textheight 22cm \textwidth 15cm \topmargin 1mm \oddsidemargin
5mm \evensidemargin 5mm

\newcommand{\beas}{\begin{eqnarray*}}
\newcommand{\eeas}{\end{eqnarray*}}
\newcommand{\bes}{\begin{equation*}}
\newcommand{\ees}{\end{equation*}}

\newcommand{\lf}{\left}
\newcommand{\ri}{\right}
\newcommand{\f}{\frac}

\def\i2           {\mbox{$\frac{i}{2}$}}

\def\al           {\alpha}

\def\bet           {\beta}

\def\lat            {\tilde{\la}}
\def\io            {{\rm Im}\, \Omega}
\def\ro            {{\rm Re}\, \Omega}

\def\del           {\delta}

\def\ep           {\epsilon}

\def\vep           {\varepsilon}
\def\et           {\eta}

\def\ga           {\gamma}

\def\Ga           {\Gamma}

\def\la           {\lambda}

\def\lah          {\hat \la}

\def\om           {\omega}

\def\ph           {\phi}

\def\ps           {\psi}

\def\rh           {\rho}

\def\si           {\sigma}
\def\Si           {\Sigma}

\def\th{\theta}

\def\hga{{\hat \gamma}}

\def\we {{\wedge}}

\begin{document}

\begin{titlepage}
\vspace*{20mm}
\begin{center}
{\LARGE \bf{{Type IIB flux compactifications on twistor bundles}}}\\ 


\vspace*{1cm}
\vspace*{20mm} \vspace*{1mm} {Ali Imaanpur}

\let\thefootnote\relax\footnotetext{Email: aimaanpu@theory.ipm.ac.ir}

\vspace*{1cm}
  
{\it Department of Physics, School of Sciences\\ 
Tarbiat Modares University, P.O.Box 14155-4838, Tehran, Iran}


\vspace*{1mm}

\vspace*{1cm}

\end{center}

\begin{abstract}
We construct a $U(1)$ bundle over $N(1,1)$, usually considered as an $SO(3)$ bundle on ${\bf CP}^2$, 
and show that type IIB supergravity can be consistently compactified over it. 
With the five form flux turned on, there is a solution for which the metric becomes Einstein. 
We further turn on 3-form fluxes and show that there is a one parameter family of solutions. 
In particular, there is a limiting solution of large 3-form fluxes for which two $U(1)$ fiber directions 
of the metric shrink to zero size. We also discuss compactifications over $N(1,1)$ to $AdS_3$. 
All solutions turn out to be non-supersymmetric.

\end{abstract}

\end{titlepage}

\section{Introduction}
Compactifying solutions of supergravity theories provide a natural way of constructing 
consistent supergravity theories in lower dimensions. Moreover, some of the solutions turn out to be 
the near horizon geometry of M(D)-branes, and thus are of significance in AdS/CFT duality.  Most of such 
solutions, however, preserve part of the supersymmetry and usually one needs to break it to construct more   
realistic models. Squashed and stretched solutions with fluxes in the compact direction, on the other hand, 
provide examples of supergravity solutions in which supersymmetry is spontaneously broken, and therefore, 
might be of interest in  building the phenomenological models in the context of AdS/CFT duality \cite{DUF2}.

Recently, we constructed  new solutions of eleven-dimensional supergravity compactifying 
it to $AdS_5$ and $AdS_2\times H^2$.  We employed canonical forms on $S^7$ to 
write consistent ans\"atze for the 4-form field strength. The twistor space construction of ${\bf CP}^3$ 
was the key for identifying the new solutions \cite{ALI}, and as we will see in this paper, this construction 
also proves useful in finding yet more solutions. 

In this note, we extend the construction of \cite{ALI} to the case of compact manifold $N(1,1)$, and use the 
twistor space language to describe it as $U(1)$ bundle over a base which itself is an $S^2$ bundle on ${\bf CP}^2$; 
the flag manifold. The first supergravity solutions of this kind were found in \cite{ROM, SOR}, and then explicit (squashed) metrics were constructed \cite{PAG}. 
In section 2, first we consider $N(1,1)$ as an $SO(3)$ bundle over ${\bf CP}^2$ and then rewrite the metric as a 
$U(1)$ bundle over the flag manifold. We then show that on this 7-dimensional manifold there exists a natural 
harmonic 2-form, the K\"ahler form of ${\bf CP}^2$, and use it to construct an 8-dimensional twistor bundle: 
a $U(1)$ bundle over $N(1,1)$. 
Interestingly, as the harmonic 2-form is anti-self-dual, the Ricci tensor of this 8-dimensional metric in a suitable 
basis is diagonal with constant components. On the other hand, in section 3 we show that on this 8-dimensional 
manifold there exists a harmonic 3-form which we use to write down an ansatz for the 5-form field strength of type IIB supergravity. In this way, we are able to reduce the field equations to a set of algebraic equations. 
Among the three solutions we obtain one is Einstein. In section 3.1, we generalize our solution by turning on 3-form fluxes, and show that there is a one parameter family of such solutions. In a limit of large 3-form fluxes two $U(1)$  fiber directions of the metric shrink to zero size. In section 3.2, we discuss the supersymmetry of the solutions and show that they break supersymmetry. Section 4 is devoted to a discussion of 
compactification on $N(1,1)$ and the supersymmetry of the solution. Conclusions and the discussion are brought in 
section 5.

\section{$U(1)$ bundles over $N(1,1)$}
$N(1,1)$ can be considered as an $SO(3)$ bundle over ${\bf CP}^2$ admitting two Einstein metrics, and hence providing 
Freund-Rubin type solutions of eleven-dimensional supergravity \cite{RUB, DUF2}. The bundle structure is very similar to that of $S^7$ where it is viewed as an $SU(2)$ bundle over $S^4$. However, $N(1,1)$ admits a 2-form, the K\"ahler form of ${\bf CP}^2$, which, as we will see, is anti-self-dual and harmonic. This allows us to construct a $U(1)$ bundle over $N(1,1)$ so that the Ricci tensor is diagonal and has constant coefficients. Therefore, with a suitable ansatz for the form fields we are able to reduce the field equations to some algebraic equations.   

\subsection{$N(1,1)$ as an $SO(3)$ bundle over ${\bf CP}^2$}
Let us start by taking the following 7-dimensional metric of $N(1,1)$ written as an 
$SO(3)$ bundle over ${\bf CP}^2$ \cite{PAG, DUF2}: 
\bea
ds^2_{N(1,1)}\!\! &=&\!\! d\mu^2 +\f{1}{4} \sin^2 \mu\, (\Sigma_1^2 +\Sigma_2^2+ \cos^2\mu\Sigma_3^2)\nn \\
\!\! &+&\!\! \la^2\lf(
(\sigma_1 -\cos{\mu}\, \Sigma_1)^2+(\sigma_2 -\cos{\mu}\, \Sigma_2)^2+(\sigma_3 -\f{1}{2}(1+\cos^2{\mu})\, \Sigma_3)^2\ri)\label{MET1}
\eea
where $\la$ is the squashing parameter. Here $0\leq \mu \leq \pi/2$, and $\Sigma_i$'s  are a set of left-invariant one-forms on $SU(2)$: 
\bea
&&\Si_1=\cos \ga\, d\al +\sin \ga \sin \al\,  d\bet\, , \nn \\
&&\Si_2=-\sin \ga\, d\al +\cos \ga \sin \al\,  d\bet\, , \  \nn \\
&& \Si_3= d\ga +\cos \al\,  d\bet \, , \nn
\eea
with $0\leq \ga \leq 4\pi ,\, 0\leq \al \leq \pi ,\, 0\leq \bet \leq 2\pi$. There is a similar expression 
for $\si_i$'s: 
\bea
&&\si_1= \sin \ph\, d\th +\sin \th \cos \ph \,  d\tau\, ,\nn \\
&&\si_2= -\cos \ph\, d\th +\sin \th \sin \ph\,  d\tau\, , \  \nn \\
&& \si_3= -d\ph +\cos \th\,  d\tau \, , \nn
\eea
where they now take value on $SO(3)$, i.e., $0\leq \tau \leq 2\pi ,\, 0\leq \th \leq \pi ,\, 0\leq \ph \leq 2\pi$. 
They satisfy the $SU(2)$ algebra; $d\Sigma_i=-\f{1}{2}\, \ep_{ijk}\, \Sigma_j \we \Sigma_k\, , \  d\sigma_i=-\f{1}{2}\, \ep_{ijk}\, 
\sigma_j \we \sigma_k\, ,\label{SIG}$ with $i,j,k,\ldots =1,2,3$. 

As in the case of $S^7$, we can see that metric (\ref{MET1}) can be 
rewritten as a $U(1)$ bundle over a base which itself is an $S^2$ bundle on ${\bf CP}^2$, the flag manifold,  \cite{GIB, FONT, ALI}:
\bea
ds^2_{N(1,1)}&=&  d\mu^2 +\f{1}{4} \sin^2 \mu\, (\Sigma_1^2 +\Sigma_2^2+ \cos^2\mu\, \Sigma_3^2)+\la^2(d\th -\sin \ph A_1+
\cos \ph A_2)^2 \nn \\
&+&\la^2\sin^2 \th\, (d\ph -\cot \th(\cos \ph A_1 +\sin \ph A_2)+A_3)^2 +\la^2(d\tau-A)^2\, , \label{NEE}
\eea 
where, 
\be
A_1=\cos{\mu}\ \Sigma_1 \, ,\ \ \ A_2=\cos{\mu}\ \Sigma_2 \, ,\ \ \ \ A_3=\f{1}{2}(1+\cos^2{\mu})\ \Sigma_3 \, ,
\ee
and,
\be
A= \cos \th \, d\ph+\sin \th(\cos \ph A_1 +\sin \ph A_2)+\cos \th A_3 \, . \label{GAU}
\ee

In the new form of the metric, (\ref{NEE}), we can further rescale the $U(1)$ fibers to $\lat$ so that 
the Ricci tensor, in a basis we introduce shortly, is still diagonal. So, let us take the metric to be
\bea
ds^2_{N(1,1)}&=& d\mu^2 +\f{1}{4} \sin^2 \mu\, (\Sigma_1^2 +\Sigma_2^2+ \cos^2\mu\Sigma_3^2)+\la^2 (d\th -\sin \ph A_1+\cos \ph A_2)^2 \nn \\
&+&\la^2 \sin^2 \th\, (d\ph -\cot \th(\cos \ph A_1 +\sin \ph A_2)+A_3)^2 +\lat^2(d\tau-A)^2\, ,\label{NE}
\eea 
and choose the following basis
\bea
&& e^0=\, d\mu\, , \ \ \ e^1=\f{1}{2} \sin \mu\, \Sigma_1\, ,\ \ \ \ e^2=\f{1}{2} \sin \mu\, \Sigma_2\, , 
\ \ \ \  e^3=\f{1}{2} \sin \mu\cos \mu\, \Sigma_3\nn \\ 
&& e^5=\la (d\th -\sin \ph A_1+\cos \ph A_2) \, ,\nn \\
&& e^6=\la \sin \th(d\ph -\cot \th(\cos \ph A_1 +\sin \ph A_2)+A_3)\, ,\nn \\
&& e^7=\lat (d\tau-A)\, . \label{VI}
\eea
In this basis the Ricci tensor is diagonal and reads
\bea
&& R_{00}=R_{11}=R_{22}=R_{33}=6-4\la^2-2{\lat^2} \, ,\nn \\
&& R_{55}=R_{66}=4\la^2+{1}/{\la^2}-{\lat^2}/{2\la^4}\, ,\ \ \ \ \ R_{77}=4\lat^2+{\lat^2}/{2\la^4}\, .\label{N11}
\eea 
For $\la^2=\lat^2=1/2$, and $\la^2=\lat^2=1/10$ the metric becomes Einstein, and thus one can get a solution of the Freund-Rubin 
type \cite{RUB}. One can also turn on the 4-form flux in the compact direction to get the Englert type solutions \cite{ROM, PAG}.

\subsection{$U(1)$ bundles over $N(1,1)$}
To start our discussion of constructing $U(1)$ bundles we need to borrow some preliminary results,   
adapted to the $N(1,1)$ case, from \cite{ALI}. Let us first introduce the following three 2-forms
\bea
R_1&=& \sin \ph (e^{01}+e^{23}) -\cos \ph(e^{02}+e^{31}) \, ,\nn \\
R_2&=& \cos \th\cos \ph (e^{01}+e^{23})+ \cos \th \sin \ph (e^{02}+e^{31})-\sin \th (e^{03}+e^{12})\, ,\nn \\
K&=&  \sin \th\cos \ph (e^{01}+e^{23})+ \sin \th \sin \ph (e^{02}+e^{31})+\cos \th (e^{03}+e^{12})\, ,\label{THREE}
\eea
with the angles and basis given in the previous subsection. These three 2-forms are orthogonal to each other, i.e.,
\be
R_1\we R_2=K\we R_1=K\we R_2=0\, .\label{R1}
\ee
With  $A$ in (\ref{GAU}) rewritten as 
\be
A=\cot \th\, \f{e^6}{\la}+\f{2\cot \mu }{\sin \th}\, (\cos \ph\, e^1 +\sin \ph\, e^2)\, ,
\ee
it is easy to prove that
\be
de^5=-e^6 \we A +2\la R_1\, , \ \ \ \ \ de^6=e^5\we A +2\la R_2\, .\label{56}
\ee
Further, if we define
\be
{\rm Re}\, \Omega =R_1\we e^5 + R_2 \we e^6 \, ,\ \ \ \ {\rm Im}\, \Omega =R_1\we e^6 - R_2 \we e^5\, ,\label{R2}
\ee
using (\ref{56}), we can see that
\bea
&&d\ro= 8\la\om_4-\f{2}{\la}\, e^{56}\we K\, ,\ \ \ \  \ d\io =0 \, ,\label{LL}
\eea
with $\om_4= e^0\we e^1\we e^2\we e^3 \, ,$ the volume element of the base, which is closed; $d\om_4 =0$. 

We can now look at an interesting feature of $N(1,1)$ as a bundle over ${\bf CP}^2$. The 
base manifold admits a closed 2-form, i.e., the K\"ahler form:
\be
J=\f{1}{4}\, da=\f{1}{4}\, d(\sin ^2\mu\, \Sigma_3)= e^{03}-e^{12}\, ,
\ee
so that $dJ=0$. Moreover, we observe that on $N(1,1)$ with metric (\ref{NE}) $J$ is also co-closed:
\be
d*_7 J=-d(J\we e^{567})=-2\la \, J\we \io \we e^7 -\lat\, J\we e^{56}\we (2K+e^{56}/\la^2)=0\, , \label{J}
\ee
where we used
\be
de^{56}=2\la\, \io \, ,\ \ \ \ de^7=-\lat F=-\lat dA=\lat( 2K + e^{56}/\la^2)\, ,\label{7}
\ee
and,
\be
J\we K =J\we \io=0\, ,\label{JK}
\ee
as $K$ and $\io$ are self-dual, whereas $J$ is anti-self-dual on ${\bf CP}^2$. All this indicates that we can use the 
corresponding $U(1)$ connection of $J$ to construct a $U(1)$ bundle over $N(1,1)$ so that its Ricci tensor is diagonal with constant coefficients. Therefore, for the metric of this 8-dimensional manifold, $M$, 
we take
\be
ds^2_8 = ds^2_{N(1,1)} +\lah^2 (dz -a)^2\, ,\label{8MET}
\ee
with $\lah$ measuring the scale of the new $U(1)$ fiber. Adding
\be
e^8=\lah (dz-a)\, ,
\ee
to the vielbein basis (\ref{VI}), the 8d Ricci tensor reads
\bea
&& R_{00}=R_{11}=R_{22}=R_{33}=6-4\la^2-2{\lat^2}-8 \lah^2 \, ,\nn \\
&& R_{55}=R_{66}=4\la^2+{1}/{\la^2}-{\lat^2}/{2\la^4}\, ,\nn \\
&& R_{77}=4\lat^2+{\lat^2}/{2\la^4}\, , \ \ \ \ R_{88}=16\lah^2\, .\label{EIN8}
\eea 
We see that as $J$ is harmonic and anti-self-dual, we do not get mixed components and the Ricci tensor 
remains diagonal.

\section{Type IIB compactifications to $AdS_2$}
We now show that the eight dimensional metric constructed above admits a harmonic 3-form, and then use this 
3-form to provide an ansatz for the five form field strength of type IIB supergravity. To begin with, we 
note that on this manifold there are generally three 4-forms which are closed and self-dual on  ${\bf CP}^2$ 
\cite{ALI}. On the other hand, since $de^8=-4\lah J$ is anti-self-dual we can write down a 5-form which is also 
closed:
\be
*_8\, \om_3 = (\al \om_4+\bet K\we e^{56}+\ga e^7\we \io )\we e^8+\xi J\we e^{567}\, ,
\ee
with $\al\, ,\bet\, ,\ga\, ,$ and $\xi$ being constant parameters. In fact, using (\ref{LL}), (\ref{7}), 
and (\ref{JK}) we can see that $d*_8 \om_3=0$. Taking the Hodge dual (with $\ep_{01235678}=1$), we have
\be
\om_3=-\al e^{567}-\bet K\we e^7+\ga \ro-\xi J\we e^8\, ,
\ee
which we also require to be closed. Using (\ref{LL}) together with
\be
dK=- \io /\la \, ,
\ee
we see that $\om_3$ is closed if
\be
\bet =2\al \la^2\, ,\ \ \ \ga=-2\al \la\lat\, ,\ \ \ \ \xi=-3\al\la^2\lat/\lah\, .\label{ABC}
\ee
Hence, on $M$ there exists a harmonic 3-form; $d\om_3=d*_8\om_3=0$.  

To discuss type IIB supergravity, we take a direct product ansatz for the metric:
\be
ds^2_{10}=ds^2_2+ds^2_8\, ,
\ee 
together with the following ansatz for the self-dual 5-form:
\be
F_5=\om_3\we \ep_2 +*_8\om_3\, ,\label{ANS5}
\ee
which then satisfies the equation of motion, $d*F_5=0$, as $\om_3$ is harmonic.  

Next, let us consider the Einstein equations. Taking the dilaton and axion to be constant, in the Einstein frame, 
they read
\bea
&&R_{MN}=\f{1}{4\cdot 4!}(F_{MPQRS}F_N^{\ PQRS}-\f{1}{10}F_{PQRSL}F^{PQRSL}\, g_{MN})\label{EIN} \\
&&+\f{e^{-\ph}}{4}(H_{MPQ}H_N^{\ PQ}-\f{1}{12}H_{PQR}H^{PQR}\, g_{MN})
+\f{e^{\ph}}{4}(F_{MPQ}F_N^{\ PQ}-\f{1}{12}F_{PQR}F^{PQR}\, g_{MN})\, .\nn
\eea
Using (\ref{EIN8}) and ansatz (\ref{ANS5}), the Einstein equations reduce to 
the following algebraic equations:
\bea
&& 6-4\la^2-2\lat^2-8 \lah^2=\al^2/4 \, ,\nn \\
&& 4\la^2+\f{1}{\la^2} -\f{\lat^2}{2\la^4}=(2\bet^2-\al^2+2\xi^2)/4\, ,\nn \\
&& 4\lat^2+\f{\lat^2}{2\la^4}=(4\ga^2-2\bet^2-\al^2+2\xi^2)/4\, ,\nn \\
&& 16\lah^2=(4\ga^2+\al^2+2\bet^2-2\xi^2)/4\, .\label{EQS}
\eea 
First we note that there is a solution for which the metric is Einstein. Plugging (\ref{ABC}) into the above 
equations, we get the following solution:
\be
\la^2=\lat^2=1/4\, ,\ \ \ \ \lah^2=3/16\, ,\ \ \ \ \al^2 =12\, ,
\ee
with the Ricci tensor along $AdS_2$:
\be
R_{\mu\nu}=-12\, g_{\mu\nu}\, .
\ee

With the help of {\em Mathematica}, we have also found two more solutions of eqs. (\ref{EQS}) for which the metric is not Einstein:
\be
\la=\lat\approx 0.4267\, ,\ \ \ \ \lah \approx 0.2661\, ,\ \ \ \ \al \approx 4.1667\, ,
\ee
with $R_{\mu\nu}\approx -14.4583\, g_{\mu\nu}$. And,
\be
\la\approx 0.5609\, ,\ \ \ \ \ \lat\approx 0.4095\, ,\ \ \ \ \lah\approx 0.4480\, ,\ \ \ \ \al \approx 3.3464\, ,
\ee
with $R_{\mu\nu}\approx -11.5538\, g_{\mu\nu}$.

\subsection{A one parameter family of solutions}
Having found a solution for which the metric is Einstein, we are interested to see whether we can 
have solutions with $H$ and $F_3$ fluxes turned on. 
For this we note that indeed there are two 3-forms which are closed:
\be
H=\zeta\, de^{78}=\zeta\, \lat(2K+e^{56}/\la^2)\we e^8 +4\zeta\, \lah e^7\we J\, ,\label{h}
\ee  
and,
\be
F_3=\eta \, \io=\eta\, de^{56}/2\la\, ,\label{f3}
\ee
with $\zeta$ and $\eta$ two constants. With the above ans\"atze for the 3-form fields, let us now turn 
to the type IIB equations of motion which, in the Einstein frame, read:
\bea
&&d* d\ph =e^{2\ph} dc\wedge *dc -\f{1}{2}e^{-\ph}H\wedge * H+\f{1}{2}e^\ph\tilde{F}_3\wedge *\tilde{F}_3 \nn \\
&&d(e^{2\ph}*dc)=-e^\ph H\wedge *\tilde{F}_3 \nn \\
&&d*(e^{-\ph}H -c e^\ph \tilde{F}_3)=  \, F_3\wedge F_5 \nn \\
&&d*(e^\ph\tilde{F}_3)=\,- H\wedge F_5  \nn \\
&& d*{\tilde F}_5= H\wedge F_3\, ,\label{SUG}
\eea
where,
\bea
&&F_3=dC_2\, ,\ \ \ F_5=dC_4\, ,\ \ \ H_3=dB\, , \nn \\
&&\tilde{F}_3=F_3-cH_3\, , \ \ \ \ \tilde{F}_5=F_5-C_2\wedge H_3\, ,\ \ \ \ \ \ *\tilde{F}_5 =\tilde{F}_5 \, .
\eea

First note that because of (\ref{R1}), (\ref{R2}), and (\ref{JK}) we have 
$H\we F_3=0$, and hence we can use the same $F_5$ as in the previous section, namely let
\be
\tilde{ F}_5=\om_3\we \ep_2 +*_8\om_3\, ,\label{f5}
\ee
so that the last equation of (\ref{SUG}) is satisfied. Taking $\ph$ to be constant, $c=0$, and the form fields 
as in (\ref{h}), (\ref{f3}), and (\ref{f5}) we can see that the rest of equations in (\ref{SUG}) are also 
satisfied if
\be
e^{\ph}\eta =\ga \zeta\, ,\ \ \ \ \ \ \ga^2=8\lah^2+2\lat^2+\lat^2/4\la^4\, .
\ee
This leaves us with four unknown coefficients to be fixed. However, when we plug these into  
Einstein equations (\ref{EIN}) they collapse into 3 equations:
\bea
&& 6-4\la^2-2\lat^2-8 \lah^2=\f{\ga^2}{16 \la^2\lat^2} +2 b^2(\lat^2+4\lah^2) \, ,\nn \\
&& 4\la^2+\f{1}{\la^2} -\f{\lat^2}{2\la^4}=\f{\ga^2}{2}\lf(\f{\la^2}{\lat^2}-\f{1}{8 \la^2\lat^2}+\f{9\la^2}{4\lah^2}\ri)
+\f{b^2\lat^2}{2\la^4}\, ,\nn \\
&& 4\lat^2+\f{\lat^2}{2\la^4}=\ga^2\lf(1-\f{1}{16 \la^2\lat^2}-\f{\la^2}{2\lat^2}+\f{9\la^2}{8\lah^2}\ri)
+b^2 \lf(8\lah^2-2\lat^2-\f{\lat^2}{4 \la^4}\ri)\label{H}\, ,
\eea 
with $b^2=e^{-\ph}\zeta^2$. So, we get a free parameter, $b$, that is not determined 
by the equations of motion. Although we have not been able to find the most general solution of (\ref{H}), 
by examining the pattern of numerical solutions generated by {\em Mathematica} we did 
derive a particular solution:
\be
\la^2=\f{1}{4}\, , \ \ \ \ \ \lat^2=\f{1}{4(1+b^2)}\, ,\ \ \ \ \lah^2=\f{3}{16(1+b^2)}\, ,
\ee
which can be checked by direct substitution in eqs. (\ref{H}). Note that for $b=0$, we get the solution in the 
previous section where we had only $F_5$ turned on. In the extreme limit $b\to \infty$, $\lat$ and $\lah$ 
go to zero and thus two $U(1)$ directions of metric (\ref{8MET}) and (\ref{NE}) shrink to zero size. 
Surprisingly, the Ricci tensor of $AdS_2$ turns out to be independent of $b$,
\be
R_{\mu\nu}=-12 \, g_{\mu\nu}\, .
\ee

\subsection{Supersymmetry}
In this section we show that the solution we found in sec. 3, where we had
turned on only the five-form flux with constant dilaton, breaks all supersymmetries.   
When the dilaton and axion are constant and there are no 3-form fluxes, the variation of 
the dilatino vanishes. However, we need to check whether the supersymmetry variation of the gravitino 
vanishes too, i.e.,
\bea
\del \ps_M &=&\nabla _M\vep+\f{i}{16\cdot 5!}\Ga^{NPQRS}\Ga_M F_{NPQRS}\, \vep =0 \, .\label{KK}
\eea
 
To study the Killing equation, (\ref{KK}),  on the direct product space $AdS_2\times M$ of sec. 3, 
let $\vep=\ep \otimes \eta$, with $\ep$ and $\eta$ the supersymmetry parameters along $AdS_2$ and $M$, respectively. 
We decompose the 10d Dirac matrices as
\bea
&& \Ga_\mu=\hga_\mu\otimes \ga_9\, ,\ \ \ \ \ \ \ \mu =0,1 \, ,\nn \\
&& \Ga_{m+1}=1\otimes \ga_m\, , \ \ \ \ \ \ m=1, \ldots, 8\, ,\nn
\eea
where $\hga_\mu$ and $\ga_m$ are the 2 and 8 dimensional Dirac matrices respectively, with $\hga_0=i\si_2$, and $\hga_1=\si_1$. 

We can see that the supersymmetry is broken by looking at the Killing equation along $AdS_2$. First, note that
\be
F_{NPQRS}\Ga^{NPQRS}=10\, F_{mnp\mu\nu}\Ga^{mnp\mu\nu} (1-\Ga_{11})\, ,
\ee
so if we choose $\Ga_{11}\vep=(\si_3\otimes \ga_9)(\ep\otimes \eta) =\vep$, with $\Ga_{11}=-\Ga_{0123456789}$, 
then we have
\bea
&&F_{NPQRS}\Ga^{NPQRS}\Ga_\mu \, \vep=20\, F_{mnp\rh\si} \Ga^{mnp\rho\si}\Ga_\mu \vep \\ \nn
&&=40\, F_{mnp01}(1\otimes \ga^{mnp})(\si_3\otimes 1)(\hga_\mu\otimes \ga_9)(\ep\otimes\eta)\\ \nn
&&=40\, F_{mnp01}(\hga_\mu \otimes \ga^{mnp})(\ep\otimes\eta)\, .
\eea
Therefore, to split  Killing equation (\ref{KK}) along the $AdS_2$ and the compact direction, we need to require
\be
F_{mnp01}\ga^{mnp}\eta=k \et\, ,\label{F}
\ee
for $k$ a constant, so that along $AdS_2$ we have
\be
\nabla_\mu\ep +40 k\, \hga_\mu \ep=0\,  .
\ee
But, since $\ga_{mnp}$ anticommutes with $\ga_9$ and since $\et$  has a definite chirality, $\ga_9\et=\et$, 
eq. (\ref{F}) can only have a zero eigenvalue, i.e., we must have $k=0$. 
On the other hand, if $k=0$, then the integrability of Killing spinor equation $\nabla_\mu \ep=0$ implies that 
the 2-dimensional Ricci tensor is vanishing which is not consistent with the $AdS_2$ factor that we obtained 
from solving the equations of motion. Therefore we conclude that the solution breaks supersymmetry. 
The above argument also applies to the solutions of sec. 3.1.

\section{Type IIB on $N(1,1)$}
Now that we have discussed the compactification of type IIB on $U(1)$ bundles over $N(1,1)$, let us look at 
the related compactification of type IIB on $N(1,1)$ itself. We study solutions with only $F_5$ flux turned on.  
So, let us  take a direct product ansatz for the metric:
\be
ds^2_{10}=ds^2_3+ds^2_{N(1,1)}\, ,
\ee 
together with $F_5$ as 
\be
F_5= \al\, (J\we \ep_3+J\we e^{567})\, ,
\ee
which is self-dual and closed because of (\ref{J}). Using this ansatz and the Ricci components of $N(1,1)$ in (\ref{N11}), 
Einstein equations (\ref{EIN}) reduce to
\bea
&& 6-4\la^2-2\lat^2=0 \, ,\nn \\
&& 4\la^2+\f{1}{\la^2}-\f{\lat^2}{2\la^4} =\al^2/2\, ,\nn \\
&& 4\lat^2+\f{\lat^2}{2\la^4}=\al^2/2\, ,\nn 
\eea 
which have just one solution;
\be
\la^2=\lat^2=1\, ,\ \ \ \ \ \ \al^2=9\, .
\ee
Note that in this solution the Ricci tensor along the base manifold, i.e., ${\bf CP}^2$, vanishes. The non-compact space is an $AdS_3$ with
\be
R_{\mu\nu}=-9/2\, g_{\mu\nu}\, .
\ee

Finally, let us discuss the supersymmetry of this solution. For $AdS_3\times N(1,1)$ compactification, 
we take the 10d Dirac matrices as
\bea
&& \Ga_\mu=\hga_\mu\otimes 1 \otimes \si_1\ \ \ \  \ \ \ \ \mu =0,1,2 \, ,\nn \\
&& \Ga_{m+2}=1\otimes \ga_m\otimes \si_2\, ,\ \ \ \ \  m=1, \ldots, 7\, ,\nn
\eea
where $\hga_\mu$ and $\ga_m$ are 2 and 8 dimensional Dirac matrices respectively, with $\hga_0=i\si_2\, , \hga_1=\si_1$, and $\hga_2=\si_3$. 
The supersymmetry parameter then decomposes 
\be
\vep=\ep \otimes \eta\otimes \left(
\begin{array}{l}
1 \\
0	
\end{array}\right)\, ,
\ee
with $\Ga_{11}=1\otimes 1\otimes\si_3$. As in the previous section, let us first look at the Killing 
equation along the $AdS_3$ factor. Note that
\bea
&& F_{NPQRS}\Ga^{NPQRS}\Ga_\mu \vep =10\, F_{mn\nu\rh\si}\Ga^{mn\nu\rh\si}\Ga_\mu(1+\Ga_{11})\vep\nn \\
&&= 5!\, F_{mn 012}(\hga_\mu \otimes \ga^{mn} \otimes 1)(\ep \otimes \eta\otimes \left(
\begin{array}{l}
1 \\
0	
\end{array}\right))\, ,
\eea
so, to split the Killing equation along $AdS_3$ and $N(1,1)$ we must have
\be
(\ga^{03}-\ga^{12})\eta =2 i\eta\, .\label{ET}
\ee
The integrability of the Killing equation along the base manifold ${\bf CP}^2$, on the other hand, requires
\bea
&&(-2 \ga^{01}+\ga^{23}-\ga^{\hat{2}\hat{3}}+\tilde{c} \ga^{\hat{1}}(\ga^{03}-\ga^{12}))\, \et=0 \, ,\nn \\ 
&&(-2 \ga^{23}+\ga^{01}-\ga^{\hat{2}\hat{3}}+\tilde{c} \ga^{\hat{1}}(\ga^{03}-\ga^{12}))\, \et=0 \, ,\nn 
\eea
with $\tilde{c}$ a constant. The above equations imply
\be
(\ga^{03}-\ga^{12})\eta =0\, ,
\ee 
which is in conflict with eq. (\ref{ET}). Therefore, we conclude that the solution breaks supersymmetry. 

\section{Conclusions}
We constructed a $U(1)$ bundle over $N(1,1)$, and showed that type IIB supergravity can be consistently 
compactified over it. The twistor space formalism was crucial in deriving the solutions, specially 
when there were 3-form fluxes turned on. This approach has earlier been used in deriving new eleven-dimensional supergravity solutions \cite{ALI}, and also in \cite{TOM} to study new solutions of massive IIA supergravity. 

We noticed that $N(1,1)$ admits a harmonic 2-form and used it to write the 
twistor bundle eight dimensional metric. With this choice of the connection, the Ricci tensor turned out to be 
diagonal with constant components. Furthermore, we saw that this eight dimensional manifold allows a harmonic 
3-form, which was then employed to write a consistent ansatz for the 5-form field strength of type IIB supergravity. 
In this way, we showed that the field equations could be reduced to a set of algebraic equations. Among the three 
solutions we found one was Einstein. The discussion became more interesting when we turned on 3-form fluxes and 
obtained a one parameter family of solutions. Amusingly, we observed that there was a limiting solution for which 
two fiber directions of the metric were shrinking to zero size, whereas the 2-dimensional cosmological constant 
turned out to be  independent of the free parameter. At the end, we further studied the related compactification over $N(1,1)$ to $AdS_3$. 

Since all the solutions we found in this paper break supersymmetry it is interesting to see  whether they are associated with some brane configurations. This would then allow us to study AdS/CFT in a non-supersymmetric 
set up.

\hspace{30mm}


\vspace{1.5mm}

\noindent

\vspace{1.5mm}

\noindent

\newpage



\end{document}